\documentclass[aps,prb,showpacs,floats,epsfig]{revtex4}
\usepackage{amssymb}
\usepackage{amsbsy}
\usepackage{amsmath}
\usepackage{epsfig}
\usepackage{amsfonts} 
\usepackage{bm}
\usepackage{graphicx}
\usepackage[breaklinks,colorlinks = true,linkcolor =
blue,urlcolor = blue,citecolor = blue,anchorcolor =
green,bookmarks=true]{hyperref}
\usepackage{mathrsfs}
\allowdisplaybreaks

\newcommand{\gae}{\lower 2pt \hbox{$\,
\buildrel{\scriptstyle >}\over {\scriptstyle \sim}\,$}}
\newcommand{\lae}{\lower 2pt \hbox{$\,
\buildrel{\scriptstyle <}\over {\scriptstyle \sim}\,$}}
\begin{document}

\title {FCI-QMC approach to the Fermi polaron}

\author{M. Kolodrubetz$^{1}$, B. K. Clark$^{1,3}$}
\affiliation{$^{1}$Department of Physics, Princeton
University, Princeton, NJ 08544, USA.}
\affiliation{$^{3}$Princeton Center for Theoretical Science, Princeton 
University, Princeton, NJ 08544, USA.}
\date{\today}

\begin{abstract} Finding the ground state of a fermionic
Hamiltonian using quantum Monte Carlo is a very difficult
problem, due to the Fermi sign problem.
While still scaling exponentially, full configuration-interaction
Monte Carlo (FCI-QMC) \cite{Booth2009_1} mitigates some of 
the exponential variance by allowing annihilation of noise -- whenever two
walkers arrive at the same configuration with opposite
signs, they are removed from the simulation. While FCI-QMC
has been quite successful for quantum chemistry problems
\cite{Booth2009_1, Cleland2010_1}, its application to
problems in condensed systems has been limited
\cite{Shepherd2011_1,Spencer2011_1}. In this paper, we apply
FCI-QMC to the Fermi polaron problem, which provides an ideal
test-bed for improving the algorithm.  In its
simplest form, FCI-QMC is unstable
for even a fairly small system sizes. However, with a 
series of algorithmic improvements, we are able to
significantly increase its effectiveness. We 
modify fixed node QMC to work in these systems,
introduce a well chosen importance sampled trial
wave function, a partial node
approximation,  and a variant of released node.  
Finally, we develop a way to perform FCI-QMC directly in the
thermodynamic limit
\end{abstract}


\maketitle

Full configuration-interaction quantum Monte Carlo (FCI-QMC)
\cite{Booth2009_1} is a method for finding the ground state
of a fermionic Hamiltonian $H$. Starting from some state
$|\psi_i\rangle$ with non-zero ground state overlap, FCI-QMC
stochastically performs imaginary time propagation
$|\psi_\beta \rangle = e^{-\beta H} |\psi_i \rangle$.
Working in a second quantized formalism, where anti-symmetry
enters in the off-diagonal components of $H$,
FCI-QMC formally yields the fermionic ground state $|\psi_0\rangle$
for sufficiently large $\beta$.

The wavefunction at any instant in imaginary time is sampled
as a set of $N_w$ walkers, where each walker is a signed
element of some many-body basis. Breaking up $e^{-\beta H}$
into $N$ small time steps, $e^{-\beta H}=e^{-\beta H / N}
e^{-\beta H / N} \cdots e^{-\beta H / N} \approx (1-\tau
H)^N$, where $\tau=\beta/N$, one applies the operator $U_1 =
1-\tau H$ stochastically \cite{Footnote_FCIFeedback}.

The fermion sign problem manifests in the negative
off-diagonal elements of $U_1$, whose sign is set by Fermi
statistics \cite{Kolodrubetz2012_1}. In general, there will
be loops such that a walker starts in some basis state
$|D\rangle$, hops around the loop via the off-diagonal
elements of $U_1$, and then returns to $|D\rangle$ with the
opposite sign. If these sign-violating loops exist for a
given Hamiltonian and basis, they are said to have a sign problem.

FCI-QMC attempts to mitigate the sign problem by
annihilation; any time two walkers end up on
$|D\rangle$ with opposite signs, they are both removed from
the simulation. It relies on efficient annihilation to
prevent walkers with the incorrect sign from propagating
as noise. In principle one can always pick a very large
$N_w$ to achieve efficient annihilation, but the necessary
number of walkers increases significantly with basis size.
The efficiency of FCI-QMC has been explored in a few
systems, including all-electron molecules
\cite{Booth2009_1}, the homogeneous electron gas \cite{Shepherd2011_1}, and
the Hubbard model \cite{Spencer2011_1}.

In this paper we develop extensions to FCI-QMC and 
apply these new developments to  the Fermi polaron
\cite{Chevy2006_1}. The polaron is 
a classic problem in condensed matter
physics, as it is one of the simplest problems that
shows strongly-interacting many body effects. We will be
particularly interested in the Fermi polaron as seen in cold
atomic gases \cite{Schirotzek2009_1}, whose ground state
properties are well understood through a combination of
analytical \cite{Chevy2006_1} and numerical
\cite{Prokofev2008_1} methods. As such, the polaron problem
serves as a nice test-bed for testing and improving the
FCI-QMC algorithm in a condensed matter setting.

In the remainder of this paper, we  introduce the
polaron problem and  discuss how we calculate
its properties by improving on the FCI-QMC algorithm.
In particular, we develop approaches which start
with a sign-free Hamiltonian and reintroduce  these
signs so as to achieve an exact answer. In addition, we 
develop a way to do FCI-QMC
in the thermodynamic limit.  

In section \ref{sec:polaronIntro}, we introduce the polaron.
Then, for a test value of $10^6$ walkers,
we find that the naive implementation
of FCI-QMC works for a basis
restricted to only a single excitation on the 
non-interacting ground state,
but fails for cases with additional excitations.  We proceed to 
describe a trial wavefunction (section \ref{sec:impsamp}) inspired by variational
solutions to the polaron problem \cite{Combescot2008_1} 
and introduce a variant of FCI-QMC which performs
importance sampling.  This increases the accessible parameter regime,
but not sufficiently.  
Next, we introduce a controlled approximation (section 
\ref{sec:partialnode}) to attenuate
the sign problem, which we call the partial node
approximation. We extrapolate observables deduced from
partial node simulations to the exact answers.  We then 
combine release node techniques (section \ref{sec:releasenode})
with the partial
node approximation as an alternative way to find the
true ground state observables.  The 
partial node energies, although extrapolable to the correct result,
are not individually variational.
At the cost of a time-step error, we show (section \ref{sec:diagdump})
how to recover the 
variational guarantee by performing a
more typical fixed-node approach.   
To accomplish this, we overcome the non-trivial obstacle
that a single basis element is connected
by the Hamiltonian to an extremely large number of other basis elements. 
We discover that all three approaches for computing the ground state
properties are consistent with each other and in agreement with other analytical
and numerical work (section \ref{sec:results}).
Finally, we note that extrapolation to the thermodynamic limit is made
difficult by the presence of shell effects.  To avoid shell effects,
we introduce an
extension of the FCI-QMC algorithm that allows for working directly
in the thermodynamic limit (section \ref{sec:tdl}). We compare and contrast our
thermodynamic limit solution with that of diagrammatic Monte
Carlo, and comment on future possibilities for utilizing our
improvements to FCI-QMC in the field of condensed matter
physics.

\section{The polaron problem}
\label{sec:polaronIntro}

The polaron problem we consider is defined for a three-dimensional system of
two fermion species, denoted spin up and spin down. 
Starting with a non-interacting Fermi sea of
spin up particles, a single spin down ``impurity"
is added that is able to interact with the sea of spin ups.
Experimentally, near a broad s-wave Feshbach resonance the
fermions interact by a short-range potential with scattering
length $a$. The momentum space Hamiltonian is\cite{Chevy2006_1}
\begin{equation}
H=\sum_{k\sigma} \epsilon_{k}
c_{k\sigma}^\dagger c_{k\sigma}+ \frac{g}{\mathcal{V}}
\sum_{kpq} c_{k+q,\uparrow}^\dagger
c_{p-q,\downarrow}^\dagger c_{p,\downarrow} c_{k,\uparrow} ~,
\end{equation}
where $\epsilon_{k}=k^2/(2m)$, $\mathcal{V}$
is the volume of the system, and $m$ is the mass of the
particles. We work in units where $\hbar=1$.

We regularize the short-range interaction by introducing a
high-momentum cutoff $\Lambda$ on the spin-ups. The interaction
strength $g$ is cutoff-dependent \cite{Mathy2011_1}:
\begin{equation}
g^{-1} = \frac{1}{8 \pi a} - \frac{\Lambda}{4 \pi^2}~.
\end{equation}
We solve this Hamiltonian for a cube of side
length $L=\mathcal{V}^{1/3}$ with periodic boundary
conditions. With $N$ spin up particles, the Fermi wavevector
and energy are $k_F=(6\pi^2 N / \mathcal{V})^{1/2}$ and
$E_F=k_F^2/(2 m)$ respectively.

In the non-interacting limit ($a=0$), the ground
state is an undressed polaron,
\begin{equation}
|D_0\rangle
\equiv |\mathrm{FS}_\uparrow , 
{\bf 0}_\downarrow \rangle~,
\end{equation}
where $| \mathrm{FS}_\uparrow , \bf{0}_\downarrow
\rangle$ denotes a Fermi sea of spin-up particles with a
single spin down at zero momentum. Upon turning
on a weak attractive interaction (small negative $a$), the
polaron remains a well-defined quasiparticle with non-zero
quasiparticle residue $Z=|\langle \psi_0 | D_0 \rangle |^2$,
where $|\psi_0\rangle$ is the ground state of the
interacting Hamiltonian.  Note that
$Z$ can be written as the ground state
expectation $\langle \psi_0 | \mathcal{P}_0 | \psi_0 \rangle$, where
$\mathcal{P}_0=|D_0\rangle \langle D_0 |$ projects
onto the undressed polaron.

The polaron problem has been most thoroughly investigated
for the case where the scattering is unitary limited, $(k_F
a)^{-1}=0$. Chevy found that the ground state for these
parameters is well-described by a variational wavefunction
that includes all single particle-hole excitations on the
spin up Fermi sea, giving a ground
state energy of $-0.6066E_{F}$ \cite{Chevy2006_1,
Combescot2008_1}. These results were later extended by
Combescot and Mora \cite{Combescot2008_1}, who introduced a
procedure for calculating the variational ground state
energy at an arbitrary number of particle-hole excitations
and proceeded to solve for the two particle-hole pair
variational energy, $-0.6156E_{F}$.

The Fermi polaron has also been approached via Monte Carlo
techniques, most notably diagrammatic Monte Carlo
\cite{Prokofev2008_1}. Monte Carlo results for the ground
state energy agree well with the variational solutions.
Given the agreement of the ground state polaron energy
between variational expansions, Monte Carlo, and experiment
\cite{Schirotzek2009_1}, the energetics of the polaron at
unitarity are a good  test case for improvements to the FCI-QMC
algorithm.

Off of unitary on the BEC side ($a^{-1} > 0$), the polaron
eventually becomes unstable to formation of a tightly-bound
molecule. This is theoretically predicted to occur as a
first order ``phase transition" \cite{Mathy2011_1} near
$(k_F a)^{-1} = 0.9$. However, experimentally the 
quasiparticle residue
vanishes at $(k_F a)^{-1} \approx 0.75$
\cite{Schirotzek2009_1}. It is believed that this difference is due to the
small but finite density of the spin down particles in the
experiment; we will address this issue in sec.
\ref{sec:results}.

\section{FCI-QMC applied to the polaron}
\label{sec:naive}

\begin{table}[ht]
\begin{tabular}{c|c|c|c|c|c|c}
$\Lambda$ & $M$ & $N$ & $1/(k_F a)$ & $\alpha$ & $\beta$ & Basis size \\
\hline \hline
20 & 1 & 33 & 0 & -0.6 & 20 & $1.09\times10^6$ \\
\hline
20 & 2 & 33 & 0 & -0.6 & 20 & $2.88\times10^{11}$ \\
\hline
20 & 3 & 33 & 0 & -0.6 & 20 & $3.28\times10^{16}$ \\
\hline
20 & 4 & 33 & 0 & -0.6 & 20 & $2.03\times10^{21}$ \\
\hline
20 & 2 & 33 & 0.5 & -1.22 & 29.6 & $2.88\times10^{11}$\\
\hline
20 & 2 & 33 & 0.9 & -2.22 & 37.3 & $2.88\times10^{11}$\\
\hline
20 & 2 & $\infty$ & 0 & -0.6 & 20 & $\infty$ \\
\end{tabular}
\caption{Basis size and variational parameters $\alpha$ and $\beta$ used
for different Hamiltonians defined by $[\Lambda, M, N, 1/(k_f a)]$,
as described in the text.}
\end{table}

Given the effectiveness of the variational expansions in
describing the polaron, we choose as our basis all
momentum space determinants with at most $M$ particle-hole
pairs dressing the spin-up Fermi sea.  The Chevy ansatz \cite{Chevy2006_1}
is then, for example, the exact solution for $M=1$.  In this basis, all
off-diagonal elements of $H$ come from the interaction and are
given by the constant value $g/\mathcal{V}$, with their sign
set by Fermi statistics.  We fix the number of
spin-up particles $N$ and momentum cutoff $\Lambda$. Note
that we eventually want to take the physical limit $N \rightarrow \infty$ and 
$\Lambda \rightarrow \infty$, for which the basis size becomes
infinite.  We 
work in the sector with zero total momentum, which is always valid
for a polaron-like quasiparticle \cite{Mathy2011_1}. $N$
will be limited to 7, 19, 27, $\ldots$ to ensure that the
spin up Fermi sea is a closed shell.

The walkers are initialized in the undressed polaron state
with positive sign,
$|\psi_i\rangle = |D_0\rangle$. Each walker is labeled by
the pair $\{ S_w, |D_w\rangle \}$, corresponding to its sign
and determinant respectively. We will write general 
determinants as
\begin{equation}
|D\rangle = |{\bf k_1},{\bf
k_2},\ldots,{\bf k_{M}}; {\bf q_1},\ldots,{\bf q_{M}}
\rangle = c_{\bf k_1}^\dagger \cdots c_{\bf k_{M}}^\dagger
c_{\bf q_1} \cdots c_{\bf q_{M}} |D_0\rangle~,
\end{equation}
where ${\bf k}$'s label the particle excitations, ${\bf
q}$'s label the holes, ${\bf k_1} < {\bf k_2} < \ldots <
{\bf k_{M}}$, and ${\bf q_1} < \ldots < {\bf q_{M}}$ for an
arbitrary but fixed momentum ordering.  The values of ${\bf k_i}$ and
${\bf q_i}$ are restricted to momenta on a discrete grid set by the 
size of the box. 
The energy metric in this basis is
\begin{equation}
\label{eq:Enaive}
E(\beta) = \frac{\langle
D_0 | H | \psi_w (\beta) \rangle } {\langle D_0 | \psi_w
(\beta) \rangle } = \frac{\sum_w S_w \langle D_0 | H | D_w
\rangle}{\sum_w S_w \langle D_0 | D_w \rangle }~,
\end{equation} 
where the zero of energy is
defined as $\langle D_0 | H | D_0 \rangle = 0$.
For sufficiently large $\beta$, the average of $E(\beta)$
converges to the fermionic ground state energy.

\begin{figure}
\includegraphics[width=6.5in]{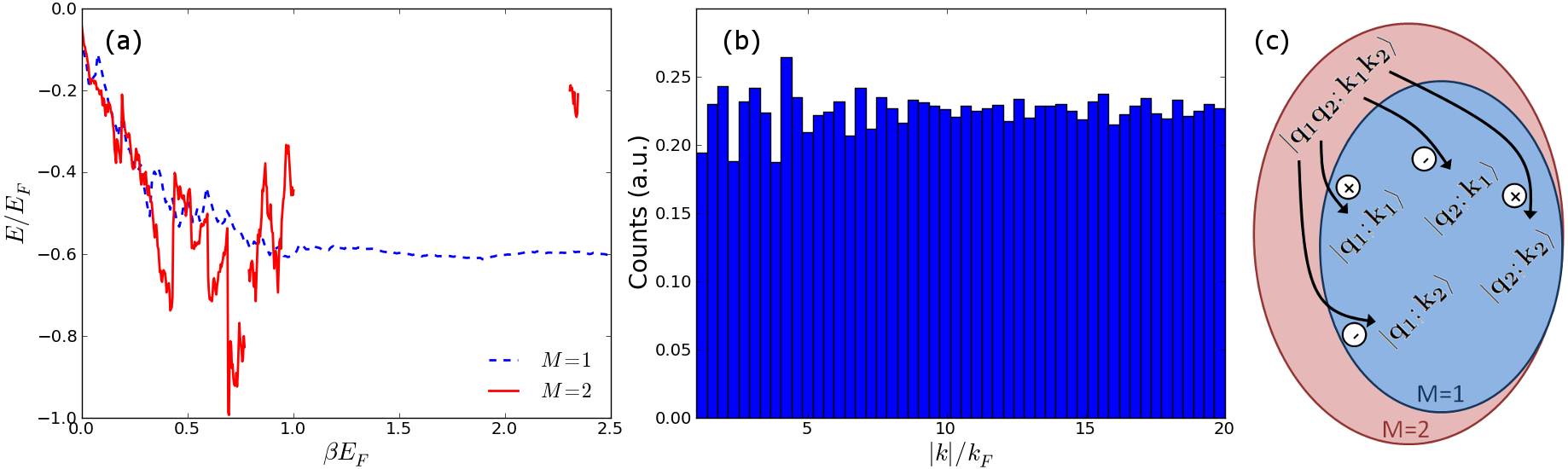}
\caption{(color online) (a) Polaron energy traces at $(k_F a)^{-1}=0.9$
for importance sampled FCI-QMC with $N=33$ spin up particles, momentum
cutoff $\Lambda / k_F=20$, $M=1$ (dashed blue), and $M=2$ (solid red).
Discontinuities in the $M=2$ data are regions where the
denominator of the energy metric (\ref{eq:Enaive}) passes
through zero. (b)
Histogram showing the momentum distribution of the particle
excitations for the $M=1$ data in part (a). Momenta are
uniformly occupied from the edge of the Fermi sea
($k=k_F$) up to the cutoff ($k=20 k_F$). (c) Graphical representation 
of sign issues for the polaron.  One can show that single particle-hole excitations
all have positive signs in the exact ground state, 
but determinants with two particle-hole pairs are connected
to them by signs that are half positive and half negative.}
\label{fig:naive}
\end{figure}

At unitarity with $N=33$ and $\Lambda=20$, FCI-QMC is able
to find the ground state energy for $M=1$, where the sign
problem is weak (fig. \ref{fig:naive}a). For $M=2$ however, the
sign problem prevents convergence with fixed number of
walkers $N_w=10^6$ (fig. \ref{fig:naive}a).
The $M=2$ sign problem
can be understood by starting from a determinant with two
particle-hole pairs, $|D_2\rangle = |{\bf k_1}, {\bf k_2};
{\bf q_1}, {\bf q_2} \rangle$. Off-diagonal matrix elements
allow hopping from $|D_2\rangle$ to four single
particle-hole determinants (see fig. \ref{fig:naive}c). Of these, two come with positive
sign, two with negative sign, and all have the same weight
$g/\mathcal{V}$. So the determinants with two particle-hole pairs rain
down nearly-random signs on the single particle-hole shell,
which one can show should all be positive in the ground state.
This results in a large sign problem unless the random signs
are efficiently annihilated.

We emphasize that our approach to the Fermi polaron differs from
previous problems where FCI-QMC has been applied because
1) we cut off the Hamiltonian after a fixed number of particle-hole
pairs, 2) we have a single particle basis that is extremely large
, and
3) almost all determinants are important.
The last fact arises because the
interaction is short-range in real space, i.e.
long-range in momentum space. 
Fig. \ref{fig:naive}b shows a
histogram of the walkers in the first particle-hole shell
for the $M=1$ ground state. The particle
excitations are evenly occupied from the Fermi sea all the
way up to the cutoff, in contrast to situations 
\cite{Booth2009_1} where  the ground state consists of
a few large-weight determinants; this makes annihilation
more difficult.

FCI-QMC for quantum chemistry problems minimizes these
issues through an intelligent choice of single particle orbitals.
However, for the Fermi polaron, there is no clear 
choice of basis that minimizes the impact of
unimportant determinants. 
In the next section we show that 
using importance sampling can enhance the quality of the algorithm.  
This not only improves the variance,
as usual in importance sampling, but also improves
the annihilation properties by favoring certain determinants.

\section{Importance Sampling and Trial Wave-Function}
\label{sec:impsamp}

Importance sampling is by now a
standard Monte Carlo technique that
has been used
to decrease statistical noise. Importance sampling
consists of sampling walker $|D\rangle$ from a
probability distribution proportional
to $|\langle D | \psi_T\rangle \langle D |  \psi_0\rangle|$ 
instead of $|\langle D |  \psi_0\rangle|$,
where $|\psi_T\rangle$ is a trial wavefunction
that we choose. This is implemented by re-weighting the
off-diagonal moves, and can be thought of as simply acting
with the (non-Hermitian) effective ``Hamiltonian"
$H_\mathrm{is}$ given by
\begin{equation}
\langle D' | H_\mathrm{is} | D \rangle = \langle D' | H | D
\rangle \frac{ \langle \psi_T | D' \rangle}
{ \langle \psi_T | D \rangle}~.
\end{equation}
The energy metric becomes
\begin{equation}
\label{eq:Eimpsamp}
E(\beta) = \frac{\sum_w S_w \langle D_0 | H | D_w \rangle
/ \langle \psi_T | D_w \rangle}
{\sum_w S_w \langle D_0 | D_w \rangle / \langle
\psi_T | D_w \rangle}~.
\end{equation}
We also calculate
quasiparticle residue through the method of mixed
estimators \cite{Ceperley1979_1},
\begin{equation}
Z\approx 2 \langle \psi_T |
\mathcal{P}_0 | \psi_0 \rangle - \langle \psi_T |
\mathcal{P}_0 | \psi_T \rangle~.
\end{equation}
As $H_\mathrm{is}$ is not Hermitian, we must be
careful to only act with it on kets.

Choosing an appropriate trial wavefunction plays a major
role in the success of importance sampling.
A naive guess, the free fermion wavefunction, 
would give $|\psi_T\rangle =
|D_0\rangle$; this would not allow sampling of any
particle-hole excitations. Therefore, we must construct
some $|\psi_T\rangle$, which in general we would like
to be as close as possible to the ground
state $|\psi_0\rangle$. 

To motivate our choice of $|\psi_T\rangle$, consider the
Schr$\ddot{\mathrm o}$dinger equation $H|\psi_0\rangle = E_0
|\psi_0 \rangle$. If $H=T+V$, where the kinetic term
$T$ is diagonal in the momentum basis, then
\begin{equation}
\langle D | \psi_0 \rangle = -\frac{1}{\langle D|\hat{T}|  D \rangle  -E_0} \sum_{D'}
\langle D | V | D' \rangle \langle D' | \psi_0 \rangle~.
\end{equation}
For $|D\rangle$ with $n$ particle-hole pairs,
the interaction term connects it to determinants $|D'
\rangle$ with $n-1$, $n$, and $n+1$ particle-hole pairs.
The strength of the interaction is $g/\mathcal{V}$ with sign
$\mathcal{S}_{D'}\equiv \mathrm{sgn}(\langle D | V | D' \rangle)$.
While these coupled linear equations are generally hard to
solve, we can make the simplifying approximation of only
considering $|D'\rangle$ with $n-1$ particle-hole pairs in
constructing $|\psi_T\rangle$.
To allow
some freedom, we will define variational
coefficients $\alpha$ and $\beta$ to replace $g$ and
$E_0$ respectively, giving
\begin{equation}
\label{eq:psit}
\langle D_n | \psi_T
\rangle = -\frac{\beta/ \mathcal{V}}{\langle D_n |\hat{T} | D_n \rangle -\alpha}
\sum_{\substack{D_{n-1}' \mathrm{s.t.}\\ \langle D_{n-1}' |
V |D_n \rangle \neq 0}} \mathcal{S}_{D_{n-1}'} \langle D_{n-1}' | \psi_T \rangle~,
\end{equation}
where $|D_n\rangle$ denotes a
determinant with $n$ particle-hole pairs. This gives a
recursive definition for $|\psi_T\rangle$ which, since normalization is
irrelevant, we seed from $\langle D_0 | \psi_T \rangle = 1$.
The parameter $\alpha$ is chosen to be approximately the energy.
Then, for each shell of $n$ particle-hole pairs, we 
calculate $Z_n=\langle \psi_T | \mathcal{P}_n | \psi_T \rangle$
using variational Monte Carlo, where $\mathcal{P}_n$ projects
onto the subspace of determinants with $n$ excitations.  $\beta$
is chosen to minimize the maximum deviation of $Z_n$ from its 
average value, $1/(M+1)$.

There is one minor issue with this $\psi_T$: 
due to the discreteness of the Hilbert space, eq.
(\ref{eq:psit}) will potentially yield $\langle D |
\psi_T\rangle = 0$ for some small set of determinants
$|D\rangle$,which become increasingly scarce as $N \to
\infty$. This prevents any weight from being put
on these determinants, introducing a bias.  To avoid this bias,
we introduce a third parameter $\gamma$ such
that if eq. (\ref{eq:psit}) yields $|\langle D | \psi_T
\rangle| < \gamma$, we set $\langle D | \psi_T \rangle \to
\gamma$. The biased simulation $\gamma=0$ 
gives a variational upper bound on the energy, 
which should be fairly close
to correct. As an example, at unitarity for the
relatively small basis with $N=7$, $\Lambda=10 k_F$, and
$M=3$, the difference in energy is $-0.6279(3)E_F$ for
$\gamma=0$ versus $-0.6310(17)E_F$ for $\gamma=10^{-4}$. We
use $\gamma=10^{-4}$ throughout the remainder of
this paper, unless otherwise specified.

\begin{figure}
\includegraphics[width=5in]{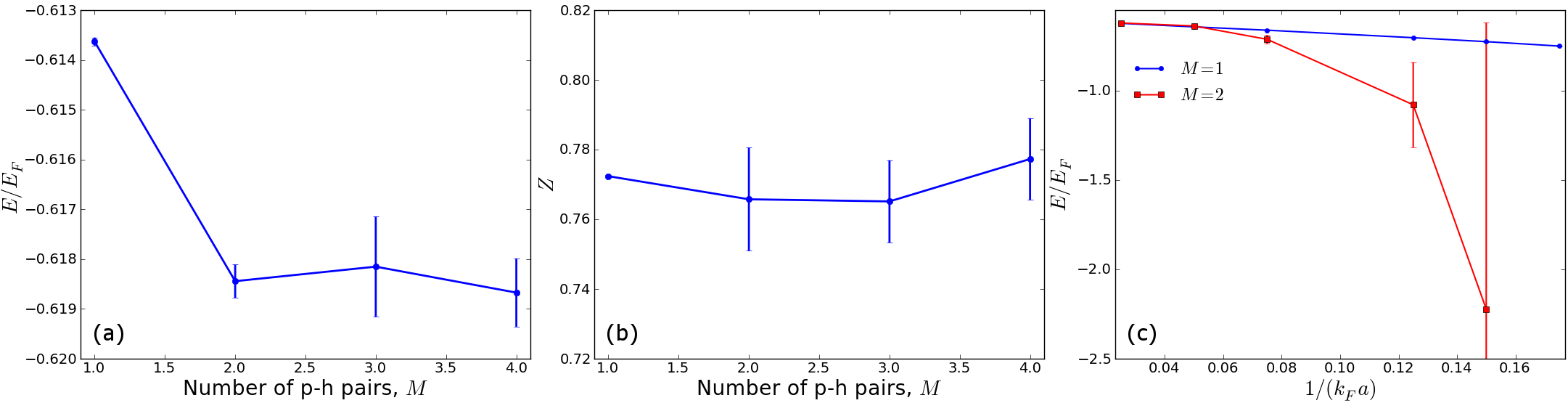}
\caption{(color online) Polaron energy (a) and quasiparticle
residue (b) for $N=33$, $(k_F a)^{-1}=0$, $\Lambda/k_F=20$,
with $M=1$ to 4 particle-hole excitations (to improve
statistics we use $\gamma=0$, so energies are a variational upper
bound). Panel (c) shows energies upon breaking off of unitarity.
Error bars are all computed for runs of  fixed imaginary time
$\Delta \beta = 35/E_F $  and therefore can be reasonably
compared with each other.
No data is shown for $(k_F a)^{-1}>0.15$ with $M=2$
because the simulation failed to converge a sign
structure in that regime.}
\label{fig:sign_breaking}
\end{figure}

Importance sampling greatly improves the effectiveness of
FCI-QMC. For example, at unitarity 
FCI-QMC with importance sampling is able to solve the
polaron ground state for
$\Lambda=20k_F$, $N=33$, and $M=4$ --
corresponding to a basis
size of $2.03\times10^{21}$ -- 
using only $10^6$ walkers.
Solutions for the polaron energy and quasiparticle
residue at unitarity are shown in fig. \ref{fig:sign_breaking}a,b.

We next push to positive values of $a^{-1}$, where
for $(k_F a)^{-1} > 0.9$ the polaron is expected to
become unstable to formation of a molecule.
As the interaction strength $1/(k_F a)$ is
increased, the polaron becomes more strongly correlated.  
The weight of the wavefunction is pushed to
higher momenta, making annihilation more difficult. As seen
in fig. \ref{fig:sign_breaking}c, for $M=2$ particle-hole pairs,
the error bar gradually increases as we push off of
unitarity. Eventually, for $1/(k_F a) > 0.15$, FCI-QMC with
importance sampling fails to find the ground state.
To proceed, we now introduce 
modifications that allow us to 
further stabilize the algorithm. 

\section{Partial node approximation}
\label{sec:partialnode}

\begin{figure}
\includegraphics[width=5in]{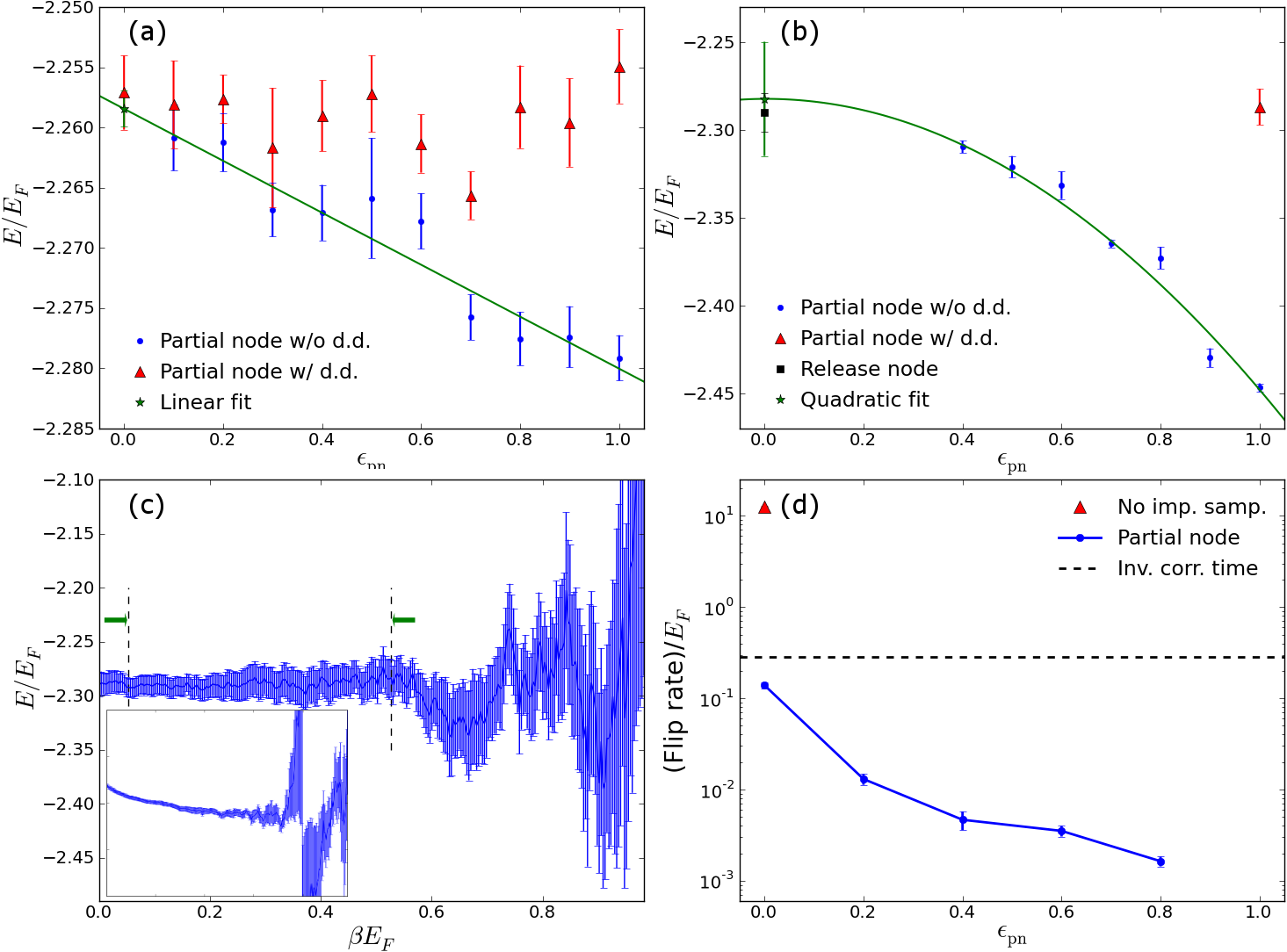}
\caption{(color online) (a) Partial node FCI-QMC energies,
both with (red triangles) and without (blue dots) diagonal dumping (see text).
Note that the $\epsilon_\mathrm{pn}=0$
result, which is computationally feasible for $M=1$, is just the solution
for the exact Hamiltonian $H$ using importance sampling.
A linear extrapolation of the partial node
energies (green star) agrees well with the $\epsilon_\mathrm{pn}=0$ solution.
(b) Partial node energies for $M=2$, where the sign problem
prevents a solution of the exact Hamiltonian. A quadratic
extrapolation agrees well with the fully-signed solution (black square),
which is obtained via release node QMC initialized from 
the ground state walkers of partial node FCI-QMC with
$\epsilon_\textrm{pn}=0.4$.
 (c) Average energy
for 100 release node traces used in panel (b). Arrows
indicate the region used in calculating statistics. Error
bars show standard error across trials. The
inset shows a similar release node trace in which the
initial condition is intentionally far from correct, to make
the decay in energy more apparent. (d) Flip rate of
the denominator of the energy metric as a function of
$\epsilon_\textrm{pn}$ (blue dots). 
 For comparison, the red triangle
shows the flip rate without importance sampling and the
black dashed line shows the inverse of the 
average the Monte Carlo correlation
time for our simulation.  All data are for
$1/(k_F a) = 0.9$, $\Lambda=20k_F$, and $N=33$. 
(a)-(c) use $10^6$ walkers, while (d) uses $10^4$.}
\label{fig:partial_node}
\end{figure}

Fixed node quantum Monte Carlo 
is a method that removes the exponential cost 
of solving a fermionic Hamiltonian by approximating the
Hamiltonian with a related one that has no sign problem.
In this section we instead find a way to extrapolate to the 
exact solution of our (restricted basis) Hamiltonian using a  fixed node-inspired
starting point. In
fixed node algorithms, such as the
lattice formulation of van Bemmel et. al. \cite{vanBemmel1994_1},
one must specify a ``correct'' sign for each
determinant. For simplicity we  choose to use
$|\psi_T\rangle$ as determining our sign structure. Given
this choice, an off-diagonal element $\langle D' |
H_\mathrm{is} | D \rangle$ will be called sign-violating if
$\langle D' | H_\mathrm{is} | D \rangle > 0$, since in this case
applying $-\tau H_\mathrm{is}$ to hop from $D$ to $D'$ would flip the
sign, violating the sign structure of $|\psi_T \rangle$. 

The fixed node Hamiltonian $H_\mathrm{fn}$ is given by
\begin{eqnarray}
\langle D' | H_\mathrm{fn} | D \rangle & = &
\left\{ \begin{array}{ll} 0 & \mbox{if sign viol. (s.v.)}\\
\langle D' | H_\mathrm{is} | D \rangle & \mbox{if not s.v. (n.s.v.)}
\end{array} \right. \mbox{, where $D\neq D'$} \nonumber \\
\langle D | H_\mathrm{fn} | D \rangle & = & \langle D |
H_\mathrm{is} | D \rangle + \sum_{D' \mbox{s.v.}} \langle D'
| H_\mathrm{is} | D \rangle~.
\label{eq:diagdump}
\end{eqnarray}
Note that the sign-violating off-diagonal matrix
elements have been removed and ``dumped" onto the diagonal.
This $H_\mathrm{fn}$ 
is guaranteed to give a variational upper bound on the
ground state energy, yielding the correct ground state
energy if $|\psi_T \rangle = |\psi_0 \rangle$. 

Lattice fixed node
was originally designed for real space lattices where there are very
few off-diagonal elements connected to any given
configuration. For our Hamiltonian, there will in general be
many sign-violating matrix elements connected to a given
determinant $|D\rangle$ (of order $>10^7$ for our parameters),
and the sums required to modify the
diagonals of $H_\mathrm{fn}$ become analytically
intractable. We therefore  initially work with a modified
Hamiltonian $H_\mathrm{fn}'$ -- which we refer as
the fixed node Hamiltonian without diagonal dumping -- given by
\begin{eqnarray}
\langle D' | H_\mathrm{fn}' | D \rangle & = &
\left\{ \begin{array}{ll} 0 & \mbox{if sign viol.}\\
\langle D' | H_\mathrm{is} | D \rangle & \mbox{if not s.v.}
\end{array} \right. \mbox{, where $D\neq D'$} \nonumber \\
\langle D | H_\mathrm{fn}' | D \rangle & = & \langle D |
H_\mathrm{is} | D \rangle~,
\label{eq:nodiagdump}
\end{eqnarray}
whose ground state
energy is no longer guaranteed to be a
variational upper bound on that of $H$. In section
\ref{sec:diagdump} we will describe
the steps needed to reintroduce the 
upper bound.

Due to the presence of annihilation in FCI-QMC,
we are able to introduce a less-severe approximation,
which we call the partial node approximation.
In partial node, we interpolate between
the exact and fixed-node Hamiltonians
\begin{equation}
H_\mathrm{pn}=\epsilon_\mathrm{pn} H_\mathrm{fn}' +
(1-\epsilon_\mathrm{pn}) H_\mathrm{is}~.
\end{equation}
Surprisingly, we find that this algorithm works up to a very
high fraction of the original sign violating terms (small
$\epsilon_\mathrm{pn}$). To obtain the exact answer, we
extrapolate to the original Hamiltonian
($\epsilon_\mathrm{pn}=0$), heuristically using a quadratic
fit (see fig. \ref{fig:partial_node}).

To quantify the improvement gained by using the partial node 
approximation, we measure the average flip rate of the sign of the 
walkers on the reference determinant $D_0$ (which typically
should contain approximately 30\% of the walkers for a given snapshot).
The flipping of this sign is a signature
of an instability of the algorithm coming from the sign problem, and the average
flip rate must be significantly below the inverse correlation time for
useful data to be garnered from the simulation.  Fig.~3d shows the average flip rate
using the partial node approximation.  Notice that $\epsilon_\textrm{pn}=0$ 
is close to the inverse correlation time, presenting a problem, but
by $\epsilon_\textrm{pn}=0.4$ the flip rate has significantly decreased. 

\section{Release Node}
\label{sec:releasenode}

In FCI-QMC without annihilation, the sign problem manifests as a variance that 
scales exponentially with both $\beta$ and the energy difference
between the fermionic and bosonic ground states
\cite{Kolodrubetz2012_1}.  This exponential scaling
in $\beta$ holds even in the presence of annihilation at finite walker number, as
annihilation can be thought of as simply reducing the fermion-boson
ground state gap. Therefore, minimizing the total $\beta$ 
needed to reach the ground state wave-function with a certain
fidelity also minimizes the effect
of the sign problem.  Thus, one can improve the convergence time of the 
importance-sampled FCI-QMC algorithm
($\epsilon_\mathrm{pn}=0$) by starting very close to the ground state.

One approach to starting near the true ground state
is to use the ground state
of $H_\textrm{pn}$
with a small value of $\epsilon_\mathrm{pn}$.
We aren't able to explicitly represent this wavefunction,
but we can prepare it stochastically via
partial node FCI-QMC. The set of $N_w$ walkers thus prepared is 
then allowed 
to evolve under the true Hamiltonian for fixed imaginary time $\beta$. 
When the wavefunction is ``released" to evolve under the
exact Hamiltonian, it quickly relaxes to the exact ground
state; this is what we call release-node FCI-QMC \cite{Ceperley1984_1}.

In release node, we use the same energy metric as in importance 
sampled FCI-QMC.  In this case, since we have now propagated with
two non-commuting Hamiltonians ($H_\mathrm{pn}$ and $H_\mathrm{is}$),
one can show that the energy metric only becomes meaningful when the 
wavefunction reaches the ground state, at which point it gives the ground
state energy.
As seen in figure \ref{fig:partial_node},
the energy converges to the
ground state energy before the errors blow up if we start close
enough to the ground state. 
The release node energies agree well with the partial node
extrapolation (fig. \ref{fig:partial_node}),
which provides a check of the validity
of our extrapolation.

\section{Diagonal dumping}
\label{sec:diagdump}

The partial node approximation that we have discussed thus
far does not give a variational upper bound on the energy,
since we removed the diagonal dumping to get the fixed
node Hamiltonian $H_\mathrm{fn}'$ (\ref{eq:nodiagdump}).
To restore the variational upper bound, we now discuss
how one can put the effects of the diagonal dumping
back into the partial node algorithm.

While the sum involved in diagonal dumping is in general
not analytically tractable, it can be done stochastically
in much the same way as off-diagonal spawning.  During the 
diagonal create/kill step \cite{Booth2009_1} we would like
to stochastically apply 
\begin{equation}
U_\mathrm{diag}=1-\tau \langle D | H_\mathrm{is} | D \rangle - 
\tau \sum_{D' \mbox{s.v.}} \langle D'
| H_\mathrm{is} | D \rangle
\end{equation}
to a walker on determinant $|D\rangle$.  The first two terms 
can be done exactly, while the sum -- denoted $\Delta K$ -- can be sampled as follows.
\begin{enumerate}
\item
From $|D\rangle$ pick another determinant $|D'\rangle$ according
to some normalized probability function $p_\mathrm{gen}(D'|D)$.
\item
If $\langle D' | H_\mathrm{is} | D \rangle$ is not sign-violating, then
$\Delta K= 0$.
\item
Otherwise, $\Delta K= \langle D' | H_\mathrm{is} | D \rangle/
p_\mathrm{gen}(D'|D)$.
\end{enumerate}

However, as $\Delta K$ is now potentially very large, multiplying
$|D\rangle$ by weight $U_\mathrm{diag}$ has the potential to be disastrous
if $U_\mathrm{diag} \ll -1$.  Therefore, we find it necessary when stochastically
dumping the diagonal to also use the approximation
\begin{equation}
1-\tau \langle D | H_\mathrm{is} | D \rangle - 
\tau \Delta K \approx e^{-\tau (\langle D
| H_\mathrm{is} | D \rangle - \Delta K)}~.
\end{equation}
This introduces a time step error which one must extrapolate to
zero; empirically a quadratic extrapolation for $\tau < 10^{-3}$
appears to be quite effective.  Partial node results with stochastic
dumping of the diagonal are shown in fig. \ref{fig:partial_node}.

\section{Results}
\label{sec:results}

\begin{figure}
\includegraphics[width=5in]{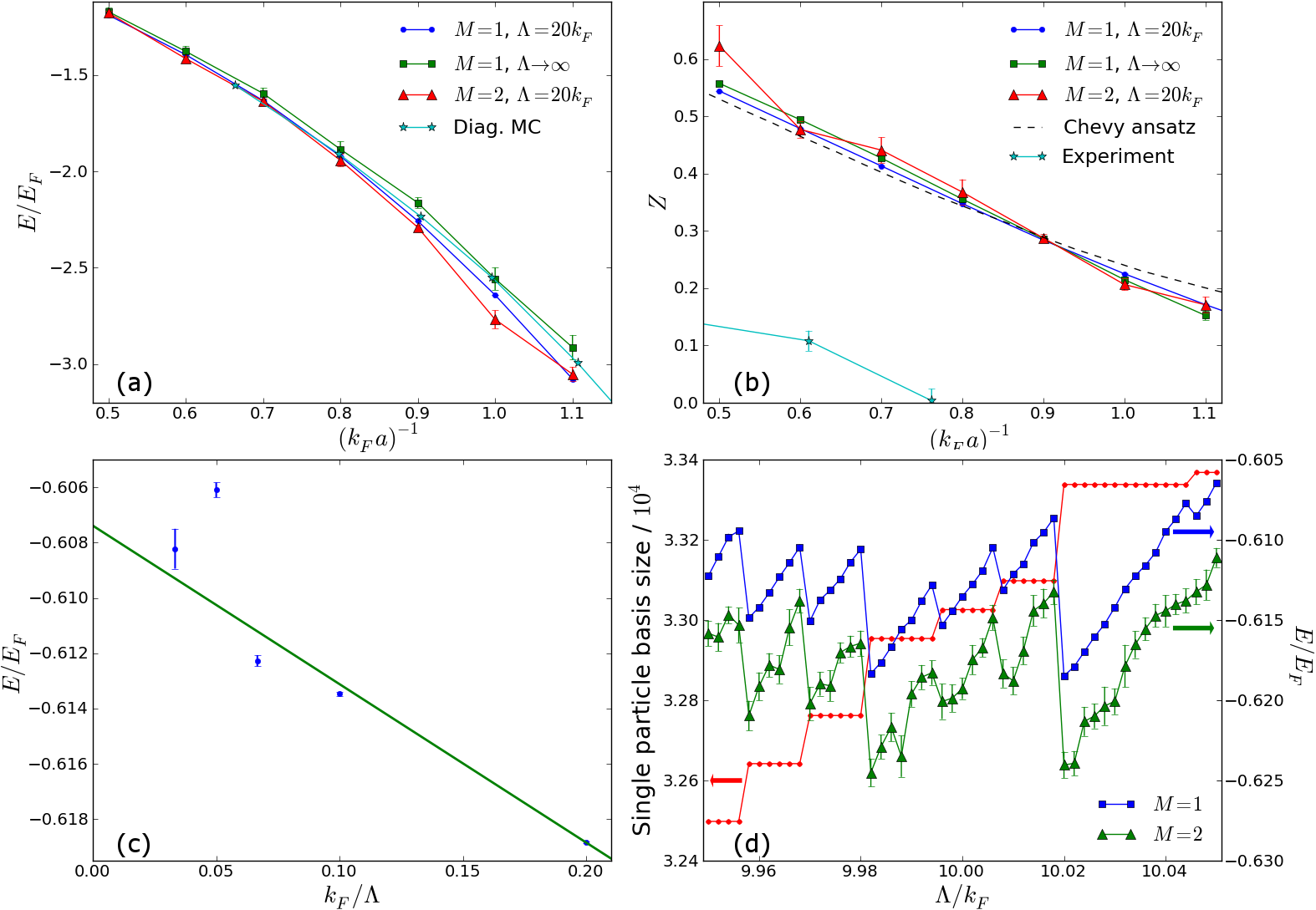}
\caption{(color online) Ground state energy (a) and
quasiparticle residue (b) for the polaron off of unitarity
at $\Lambda=20k_F$ and $N=33$. Energies are compared against
the $N\to\infty$, $\Lambda\to\infty$ results from
diagrammatic MC \cite{Prokofev2008_1}, while $Z$ is compared to
experimental measurements \cite{Schirotzek2009_1} and the
analytical variational result at $M=1$ \cite{Chevy2006_1}.
Deviations of $Z$ from the analytic results
likely come from a combination of  finite size effects and error in the mixed
estimator. (c) and
(d) show finite size effects seen in attempting to
extrapolate the energy in $1/\Lambda$, with $(k_F
a)^{-1}=0$, $N=33$, $M=1$(c) and $2$ (c,d). In (d), we
compare energy against single particle basis size, i.e.
the number of available spin up momenta $k_F<k<\Lambda$.}
\label{fig:results}
\end{figure}

Using partial node FCI-QMC followed by release node, we are
able to determine values for the energy and quasiparticle
residue of the Fermi-polaron on the BEC side of the
interaction, which are shown in fig. \ref{fig:results} for
$\Lambda=20k_F$ and $N=33$. The energies are
compared to polaron and molecule energies from diagrammatic
Monte Carlo \cite{Prokofev2008_1}. Our results match well
for small values of $1/(k_F a)$, while for larger values of
$1/(k_F a)$ our energies deviate from the diagrammatic
results. This is a consequence of working at fixed
$\Lambda=20k_F$, as opposed to diagrammatic Monte Carlo,
which is done in the limit $\Lambda \to \infty$.
Our results match with the diagrammatic
solution after extrapolating to $\Lambda=\infty$.

Figure \ref{fig:results}b shows the quasiparticle residue
$Z$ compared to experimental results
\cite{Schirotzek2009_1}, where there is a small but finite
density of spin-down atoms. We find that, as with the
energy, an $M=1$ variational expansion
provides a good estimate of $Z$. Therefore, as noted
elsewhere\cite{Schirotzek2009_1}, our theoretical model
differs significantly from the experiment, possibly as a
result of the finite spin-down density used experimentally.

In extrapolating our results to the physical
limit, $\Lambda \to \infty$, we encountered
an unexpected problem. Using a linear fit over a wide range
of $1/\Lambda$ (fig. \ref{fig:results}a), we find that
individual data points have error well outside the line,
with no discernible pattern. Zooming into a very small
region of $\Lambda$ we discovered the reason behind this:
small fractional increases in basis size as new shells
become available cause large jumps in the ground state energy. A
similar issue occurs when attempting to extrapolate in
particle number $N$.  We nevertheless attempt such an 
extrapolation in fig. \ref{fig:results} a and b.  As a result, the size
of the error bars reflects not the accuracy of the data
points at individual values of $\Lambda$, but rather these
inherent shell effects.

Shell effects will vanish in the limit $N\to\infty$. 
Therefore, we next address the possibility of
extending FCI-QMC
into the thermodynamic limit.

\section{FCI-QMC in the thermodynamic limit}
\label{sec:tdl}

\begin{figure}
\includegraphics[width=5in]{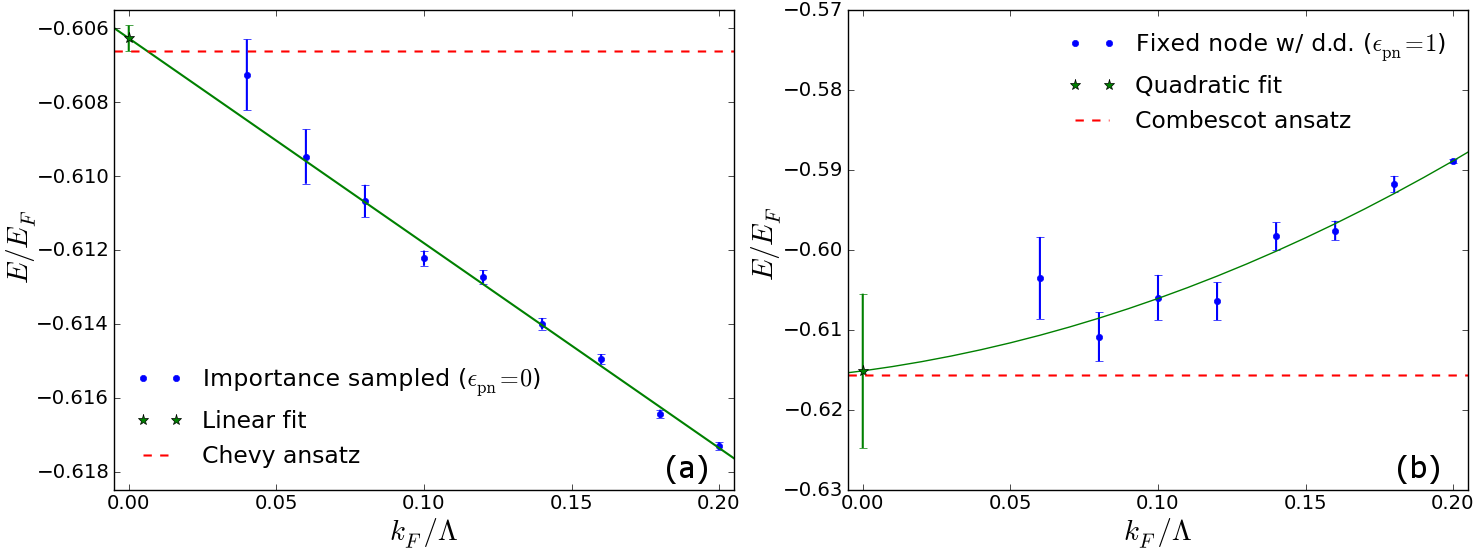}
\caption{(color online) Ground state polaron energy in the
TDL (a) with $M=1$, $(k_F a)^{-1}=0$ as a function of $\Lambda$
(blue dots). A linear extrapolation in $1/\Lambda$ (green star)
agrees with the exact ground state energy with $M=1$,
calculated by Chevy's variational ansatz 
\cite{Chevy2006_1}. (b) Fixed node
($\epsilon_\mathrm{pn}=1$) energies for $M=2$ compared
against the variational ansatz \cite{Combescot2008_1}.}
\label{fig:tdl}
\end{figure}

All the quantum Monte Carlo simulations described so far have
been done for a finite number of particles.  In this section, we 
describe how to modify FCI-QMC to work directly in the 
thermodynamic limit (TDL, $N\to\infty$).  In the thermodynamic limit, the accessible
momenta span a continuous set of k-points instead of 
being limited to a discrete grid.    Additionally, it is important
to work in a representation where instead of enumerating the 
momenta for all $N=\infty$ particles,  we instead store
only their excitations above a known state, in this case $|D_0\rangle$.
Crucially, in order for the
spectrum of the Hamiltonian to remain bounded, 
allowing application of $1-\tau H$ instead of $e^{-\tau H}$, 
the TDL only works with a cutoff $M$ on the number of excitations
allowed.  We note that there are modifications to FCI-QMC that can 
allow it to work in continuous time, i.e. allow for application of
$e^{-\tau H}$, which we discuss in appendix \ref{sec:continuoustime}. 
This requirement is similar in spirit
to the diagrammatic Monte Carlo method of imposing a cutoff
on diagram order, in an attempt to avoid a divergence.    

As momentum space is no longer discretized in the TDL, one
might assume that annihilation is no longer possible, and therefore
that the algorithm is doomed to fail.  However, there is one key
exception: due to the discrete choice of $|D_0\rangle$,
annihilation can still occur at that one determinant.

Given this, the TDL algorithm is in practice nearly identical to 
the algorithm with finite $N$ -- in fact, one can think of the TDL
algorithm as just the limit of the finite algorithm for larger and
larger $N$.  We have carefully, though implicitly, defined 
$|\psi_T\rangle$ such that 
all factors of $\mathcal{V}$ cancel out when determining
relevant quantities,
such as spawning probability or energy.  This is
important because taking the limit $N\to \infty$ while at 
fixed spin-up density (constant $k_F$) means taking 
$\mathcal{V}\to\infty$ as well.

The TDL algorithm is remarkably effective at finding the ground state
with $M=1$ at unitarity, where the sign problem is weak.  These
energies at various values of $\Lambda$ are shown in fig. \ref{fig:tdl}a,
where an extrapolation to $\Lambda=\infty$ is much smoother than
for finite $N$ due to the absence of shell effects.

For the sign-problem-heavy $M=2$ case, we instead show
data in the sign-free fixed node limit
($\epsilon_\mathrm{pn}=1$), using the diagonal dumping
method described in sec. \ref{sec:diagdump}.  This gives
a good energy for the $M=2$ polaron, which is a variational
upper bound on the known energy of $-0.6156E_F$ within
error bars.  We believe this latter approach of combining fixed node with 
working directly in the thermodynamic limit
will have wide applicability even for the approximate
calculations that currently dominate the fermion QMC literature.

\section{Discussion}
\label{sec:discussion}

We have shown that the FCI-QMC algorithm can solve the Fermi
polaron problem, after improving the algorithm with a smart
importance sampled wave function, the introduction of partial and 
fixed node approximations, and the utilization of release node methods.  

We believe that our work demonstrates two main points that should
be useful as FCI-QMC and its variants are applied to future problems in
condensed matter physics.  First, for strongly correlated condensed systems 
where many determinants are occupied in any single particle
basis, we have demonstrated that physical insight -- in the form of
a good choice of trial wavefunction -- can significantly improve the behavior
of the FCI-QMC algorithm.  Combining the improved statistics of 
importance sampling with sign-attenuating approximations such as
partial node, we showed a significant increase in the effectiveness of
FCI-QMC in solving the polaron problem.  Finally, we showed that these
methods can be made exact, up to statistical noise, via extrapolation
or release node QMC.


Second, we have shown that under certain conditions the FCI-QMC
algorithm can be extended to work with systems in the thermodynamic
limit.  Furthermore, we have introduced a fixed node algorithm in
this limit, which has historically been an important tool for
solving fermionic systems with QMC.
We anticipate that these new developments will open a variety of problems
in condensed matter physics to be approached using these new developments.  

\subsection{Acknowledgments}
We would like to thank David Huse and Charles Mathy for valuable
discussions.  This work was supported in part by ARO Award
W911NF-07-1-0464 with funds from the DARPA OLE Program. Some of the
computation was performed using the Extreme Science and Engineering
Discovery Environment (XSEDE), which is supported by National Science
Foundation grant number OCI-1053575.  Additional computational work was done
on the Feynman cluster at Princeton. 

\bibliography{C:/Users/mkolodru/Documents/Lab/References/References}

\appendix
\section{Continuous-time algorithm for FCI-QMC}
\label{sec:continuoustime}

In this appendix, we introduce a continuous-time FCI-QMC algorithm that has no fixed time step.
We note that, while we have not yet implemented this algorithm, the number of off-diagonal terms that 
must be sampled to remove the penalty method errors (discussed below) is likely to
make the algorithm significantly slower than the finite time step algorithms described earlier.

Our continuous-time formulation of FCI-QMC is loosely based on a continuous-time
lattice algorithm found elsewhere \cite{Sorella2011_1}. 
We will begin by introducing the algorithm for an arbitrary Hamiltonian $H$, which in general can be some non-Hermitian effective Hamiltonian, as found in importance sampling.  We will start by assuming that all off-diagonal sums can be computed analytically, and later generalize this to the case where certain sums must be done stochastically.  We generalize the existing algorithm to allow for situations where $H$ has a sign problem, so off-diagonal elements of $H$ will not be required to be negative.

We would like to apply the propagator $U_{\beta_A}=e^{-\beta_A H}$,
where $\beta_A$ is now some fixed imaginary time.  We
refer to $\beta_A$ as the annihilation time, and think of applying $U_{\beta_A}$
to each walker before performing annihilation.  Breaking this up into small time intervals
$\tau$, it becomes $U_{\beta_A}=(1-\tau H) (1-\tau H) \cdots (1-\tau H)$.  To perform
continuous time QMC, we would like to take the $\tau \to 0$ limit of this expression.

Consider applying $U_\tau = 1-\tau H$ stochastically to some determinant $|D\rangle$
in the limit $\tau \to 0$.  We define on-diagonal component $K_L (D)$ and off-diagonal
sum $V_L (D)$, where 
\begin{eqnarray}
K_L (D) & = & \langle D | H | D \rangle \\
V_L (D) & = & \sum_{D' \neq D} \left| \langle D' | H | D \rangle \right|~.
\nonumber
\end{eqnarray}
We can break $U_1$ up as
\begin{eqnarray}
\nonumber
U_1=\frac{1-\tau K}{1-p_s} (1-p_s) + \frac{-\tau V}{p_s}
p_s & = & K_1 (1-p_s) + V_1 p_s, \mbox{ where }\\
K_1 \equiv \frac{1-\tau K}{1-p_s} & \mbox{ and } &
V_1 \equiv \frac{-\tau V}{p_s}
\end{eqnarray}
Then we stochastically apply $K_1$ with probability $1-p_s$ or $V_1$ with probability $p_s$.

Furthermore, we would like to choose $p_s$ such that $V_1$ simply corresponds to
deterministically moving to $|D'\rangle$ with probability proportional to
$|\langle D' | H | D \rangle|$.  Thus, 
\begin{equation}
\sum_{D' \neq D} |\langle D' | V_1 | D \rangle|  =  1 \Longrightarrow p_s  =  \tau V_L (D)~.
\end{equation}

So the algorithm proceeds as follows: start from some determinant $|D\rangle$, and either
apply $K_1$ with probability $1-p_s$ or $V_1$ with probability $p_s$.  $V_1$ corresponds
to hopping to a new determinant.  $K_1$ simply multiplies $|D\rangle$ by a weight
\begin{equation}
W_1=\frac{1-\tau K_L(D)}{1-p_s} = \frac{1-\tau K_L (D)}{1-\tau V_L(D)}~.
\end{equation}
$V_1$ is chosen for the first time at step $N$ with probability
$P(N | D) = (1-p_s(D))^N$.  If $N=\beta_S / \tau$, time
$\beta_S$ prior to spawning will come from the probability distribution
\begin{equation}
P_s(\beta_S) = \left( 1-\tau V_L \right)^{\beta_S / \tau}
\stackrel{\tau \to 0}{\longrightarrow} e^{-\beta_S V_L(D)}.
\end{equation}
Given a choice of $\beta_S$ from this distribution, the walker will also pick up a total
weight $W$ during the $N-1$ non-spawning steps, where
\begin{equation}
W(\beta_S) = W_1^{\beta_S/\tau - 1} \stackrel{\tau \to 0}
{\longrightarrow} e^{-\beta_S [K_L(D) - V_L(D)]}
\end{equation}
Note that, for a non-sign-violating Hamiltonian, the term in the exponent
of $W(\beta_S)$ is just the local energy.

Therefore, for each walker we stochastically propagate for a time $\beta_A$,
during which it will both hop and pick up weight; this can be done to all the walkers
in parallel.  Finally, we take all the walkers, perform annihilation, 
measure observables, and repeat.
For Hamiltonians with relatively few off-diagonal terms, i.e. the real-space
Hubbard model, this is the complete algorithm.  For much larger off-diagonal sums,
things become more complicated.

The remainder of this discussion will describe our algorithm for Hamiltonians
with large or infinite number of off-diagonal terms in the sums.  
There are two sums (integrals in the TDL)
that we now perform stochastically: the diagonal dumping 
\begin{equation}
\Delta K_L(D)=\epsilon_\mathrm{pn} \displaystyle \sum_{D' \mathrm{s.v.}} |\langle D' | H_\mathrm{is} | D \rangle| 
\end{equation}
and the local potential energy sum
\begin{equation}
V_L (D)= \displaystyle \sum_{D' \mathrm{n.s.v.}} |\langle D' | H_\mathrm{is} | D \rangle|
+(1-\epsilon_\mathrm{pn}) \displaystyle \sum_{D' \mathrm{s.v.}} |\langle D' | H_\mathrm{is} | D \rangle|~.
\end{equation}

The sums themselves are fairly straightforward.  Using the same method 
as spawning in FCI-QMC, we have a method for generating determinant $|D'\rangle$ connected by an
off-diagonal component of $H$ to the starting determinant $|D\rangle$.
If the probability to generate $|D'\rangle$ is $p_\mathrm{gen} (D' | D)$,
then $\Delta K_L$ for example is just the expectation of the observable
\begin{equation}
\mathcal{O}_{\Delta K_L} = \epsilon_\mathrm{pn} \frac{\eta_{D D'}
\langle D' | H | D \rangle}{p_\mathrm{gen} (D' | D)} \mbox{ where } 
\eta_{D D'} = \left\{ \begin{array}{ll}
0 & \mbox{if not sign viol.}\\
1 & \mbox{if sign viol.} \end{array} \right.
\end{equation}
A similar observable $\mathcal{O}_{V_L}$ can be defined for computing $V_L(D)$.
Assume that we have sampled a total of $N_\mathrm{sum}$ determinants $D'$ to
simultaneously calculate $\Delta K_L$ and $V_L$ with some mean
$\overline{\mathcal{O}}$ and standard error $\sigma(\mathcal{O})$ for each observable.

Formally, we could perform this sampling in the limit $N_\mathrm{sum} \to \infty$,
determine $\Delta K_L$ and $V_L$ exactly, and simply use them in the earlier procedure.
However, for finite $N_\mathrm{sum}$, we want to utilize a variant of the penalty method
to minimize the bias due to statistical uncertainty.  Consider first the simpler case of applying
the weight $W=e^{-\beta_S \Delta K_L}$.  Assume that the actual value of $\Delta K_L$
is drawn from a Gaussian distribution with mean $\mu$ and width $\sigma$, which are given
by the expectation value and standard error of $\mathcal{O}_{\Delta K_L}$.  Then the weight
we should apply is simply
\begin{equation}
W_\mathrm{stoch} (\Delta K_L) \equiv \langle W \rangle = \int_{-\infty}^\infty e^{-\beta_S \Delta K_L} 
G_{\mu,\sigma}(\Delta K_L) d(\Delta K_L) = 
e^{-\beta_S (\mu - \beta_S \sigma^2/2)}
\end{equation}

A more complicated question is how to sample $\beta_S$ in an unbiased way.  Again, assume
that $V_L$ is drawn from a Gaussian $G_{\mu,\sigma}(V_L)$.  Furthermore, assume that we
have chosen $N_\mathrm{sum}$ large enough that we don't have to worry about the negative
$V_L$ tail of the Gaussian, i.e. $\mu \gg \sigma$.  We want to sample from the full distribution
\begin{equation}
p(\beta_S) = \int_{0}^\infty p(\beta_S | V_L) G_{\mu \sigma}(V_L) dV_L
\end{equation}
We can expand to second order in $V_L-\mu$, ignoring the first order term
(whose integral vanishes).  Then
\begin{eqnarray}
p(\beta_S) & \propto & \frac{e^{-\beta_S \mu}}{\mu}\int_{-\infty}^\infty
\left[  1+ \left( \frac{V_L-\mu}{\mu} \right)^2
(1+\beta \mu + \beta^2 \mu^2/2 ) + 
O\left( \left( \frac{V_L-\mu}{mu}\right)^4 \right) \right]
G_{\mu \sigma}(V_L) dV_L\\
& \approx & \frac{e^{-\beta_S \mu}}{\mu}\left[ 1 + \left( \frac{\sigma}{\mu} \right)^2
(1+\beta \mu + \beta^2 \mu^2/2 ) \right]
\end{eqnarray}
Normalizing and integrating this result, we
find that the normalized cumulative distribution function is 
\begin{eqnarray}
P(x= \mu \beta_S, s=\sigma/\mu) = 1+\frac{(\sinh x - \cosh x)
(2+s^2 (x^2+4x+6)}{6s^2+2}
\end{eqnarray}
We can then sample from this distribution by picking $z\in [0,1]$ at random,
then finding the value of $\beta_S$ where $P(x,s)=z$.

In summary, here is our algorithm for continuous partial node FCI-QMC:
\begin{enumerate}
\item
Start with $N_w$ walkers, each with weight $W_w=1$, sign $S_w$,
and determinant $|D_w\rangle$.  At the first step, all walkers are initialized
in $|D_0\rangle$ with positive sign.  At this point, there should be no walkers
with the same determinant but opposite sign.
\item
Propagate each walker independently by $e^{-\beta_A (H-S)}$ as follows.
At the start of this portion, define a variable $\beta_w$
for each walker, and initialize $\beta_w=0$.
\begin{enumerate}
\item
For a walker in determinant $|D\rangle$, sample off-diagonal elements
$|D'\rangle$ for $N_\mathrm{sum}$ steps to get $\overline{V_L(D)}$,
$\sigma [V_L(D)]$, $\overline{\Delta K_L(D)}$, and $\sigma [\Delta K_L(D)]$.
\item
Sample $\beta_S$ from $P(x,s)$ as described above, where
$x=\beta_S / \overline{V_L(D)}$ and $s=\sigma [V_L(D)] / \overline{V_L(D)}$.
\item
If $\beta_w+\beta_S < \beta_A$, then move the walker to a new determinant
$|D'\rangle$.  To sample with weight proportional to $|\langle D' | H | D \rangle|$,
run a Metropolis algorithm for $N_\mathrm{met}$ steps.  The sign of $|D\rangle$
is (not) flipped when the matrix element $\langle D' | H | D \rangle$ is positive (negative).
\item
Multiply the weight by $e^{-\beta_S(\langle D | H |D \rangle - S)}$, with the
stochastic portion of $e^{-\beta_S \langle D | H |D \rangle }$ re-weighted
via the penalty method as described above.  If $\beta_w+\beta_S > \beta_A$,
then use $\beta_A-\beta_w$ in place of $\beta_S$ in these formulas.
\item
Increment $\beta_w$ by $\beta_S$.  If $\beta_w< \beta_A$, repeat from part (a).
\end{enumerate}
\item
Annihilate the walkers, keeping track of their weights.  For example, if five walkers
are all on determinant $D$ with weights $W_1$ to $W_5$ and signs $S_1$ to $S_5$,
let $\widetilde{W} = \sum_i W_i S_i$.  Then the five walkers are replaced by a single
walker with weight $|\widetilde{W}|$ and sign $\mathrm{sgn}(\widetilde{W})$.
\item 
Resample the walkers.  Replace a single walker with weight $W$ by $\lfloor W \rfloor$
walkers of weight 1.  Add another walker of weight 1 with probability $W-\lfloor W \rfloor$.
\item
Measure observables.
\item
Adjust $S$ using standard feedback protocols \cite{Booth2009_1} as desired.
Then repeat from step 1.
\end{enumerate}

\end{document}